\documentclass[journal,10pt,draftclsnofoot,onecolumn]{IEEEtran}

\usepackage{graphicx}
\usepackage{caption}
\usepackage{subcaption}
\usepackage{float}
\usepackage{epstopdf}
\usepackage{pdfsync}
\usepackage{boxedminipage}
\usepackage{amsmath}
\usepackage{amssymb}
\usepackage{amsthm}
\usepackage{amsfonts}
\usepackage{comment}
\usepackage{cite}
\usepackage{lmodern}
\usepackage{colortbl}
\usepackage{multicol}
\usepackage{multirow}
\usepackage{epsfig}
\usepackage{psfrag}
\usepackage{url,hyperref}
\usepackage[all]{xy}
\usepackage{amsmath}
\usepackage{amsthm}
\usepackage{eucal}
\usepackage{amssymb}
\usepackage{mathrsfs}

\begin{document}

\title{Highly Scalable Neuromorphic Hardware with 1-bit Stochastic nano-Synapses}

\author{\authorblockN{Omid Kavehei and Efstratios Skafidas}
\authorblockA{Centre for Neural Engineering \\
The University of Melbourne, Victoria 3010, Australia \\
Email: \{omid.kavehei,sskaf\}@unimelb.edu.au}}

\maketitle

\begin{abstract}
Thermodynamic-driven filament formation in redox-based resistive memory and the impact of thermal fluctuations on switching probability of emerging magnetic switches are probabilistic phenomena in nature, and thus, processes of binary switching in these nonvolatile memories are stochastic and vary from switching cycle-to-switching cycle, in the same device, and from device-to-device, hence, they provide a rich in-situ spatiotemporal stochastic characteristic. This work presents a highly scalable neuromorphic hardware based on crossbar array of 1-bit resistive crosspoints as distributed stochastic synapses. The network shows a robust performance in emulating selectivity of synaptic potentials in neurons of primary visual cortex to the orientation of a visual image. The proposed model could be configured to accept a wide range of nanodevices. 

\end{abstract}

\IEEEpeerreviewmaketitle

\section{Introduction}

Cognitive computing with nanoelectronics is an emerging field of research that aims to fill the gap between CPU\rq{}s performance and mammalian brains. CPUs outperform human brain in almost all tasks that involve deterministic computation. They, however, lack a very important feature, which is the ability to learn and work with unreliable building blocks. This unique characteristic of mammalian brains has attracted attention of scientists from different fields including electronic engineering. In the late 1980s, Caver Mead, and later Eric Vittoz, envisioned the use of very-large-scale integration (VLSI) systems to mimic neuro-biological architectures of nervous system \cite{mead1989analog,vittoz1990analog}.

It is believed that the hippocampus of a mammalian brain is responsible for memory and learning. It consists of neurons and chemical synapses \cite{lim2013short}. On the contrary to the common belief that neural systems, including hippocampus, are analog systems, there are several neurobiological evidence for existence of discrete changes of synaptic strength at least in some subfields in hippocampus \cite{connor2005graded}. There are also a number of research that suggest efficient learning with digital synapses \cite{fusi2002hebbian,senn2005convergence}. Therefore, even though an analog approach promises more power efficiency, a digital implementation improves design simplicity, it does not have limited scalability of analog implementations, it is less costly, small, and by far more integrable, hence, practically the best way, if not the only way, to implement highly scalable neuromorphic systems \cite{seo201145nm}. For instance, analog designs allow low-voltage operation, which then requires larger sensing circuitry and also does not necessarily support reasonable throughputs (spikes/s). It is expected that massively parallel, reliable, scalable, and potentially energy efficient cognitive systems could be implemented using 1-bit stochastic synapses. 

A deterministic approach, has its focus on `more number of bits per synapse\rq{} to allow a more precise implementation of a learning rule, and generally suffers multiple data conversion (ADCs and DACs). On the other side, emerging artificial synapses, such as RRAM (resitive RAM), CBRAM (conductive bridge RAM), PCM (phase change memory) and atomic switches, in their analog operating mode, have issues like (i) complicated programming pulse-schemes to achieve more intermediate states, which sometimes  only exists in one switching direction (either OFF-ON or ON-OFF), and (ii) the problem of resistance drift or state retention \cite{kuzum2013synaptic,suri2013impact}. Technologies like STT-MRAM (spin-transfer torque magnetoresistive RAM) and MeRAM (magnetoelectric RAM) also show existence of multi-stable states, but their low signal-to-noise (SNR) properties \cite{wang2013low} suggests them impractical for conventional analog neuromorphic computing. A workaround to these problems is to make the artificial synaptic transmission a probabilistic process, which can be achieved either in-situ (randomness within the device) or ex-situ (external random number generator) \cite{walmsley1987probabilistic,suri2012cbram,kuzum2013synaptic,siddharth2013stochastic,yu2013low}. 

This work uses the in-situ stochastic switching characteristic of RRAM devices to present a technology independent approach to implement a highly scalable neuromorphic hardware that is capable of emulating sensitivity of synaptic potentials in neurons of primary visual cortex to the orientation of a visual image. Considering the underlying physical mechanisms of binary switching in RRAM, CBRAM and PCM it can be envisioned that their yield-loss and variation remain substantially larger than their CMOS counterparts. Therefore, a probabilistic approach is one of the most practical and highly scalable designs for neuromorphic hardware and logic operations with nanoelectronics.

\section{Stochastic nano-Synapse}\label{sec:stoch-synapse}

\subsection{Stochastic finite state machines}\label{subsec:stoch-fsm}
Stochastic binary switching is common between the emerging memory technologies. This behavior normally occurs under certain condition, which is called ``weak'' programming. Although, the weak programming condition is different across technologies, a technology independent approach is presented to pave the way for further unification of system-level behavior based on above mentioned technologies. 

\begin{figure}[!ht]
\centering
  \includegraphics[width=0.45\textwidth]{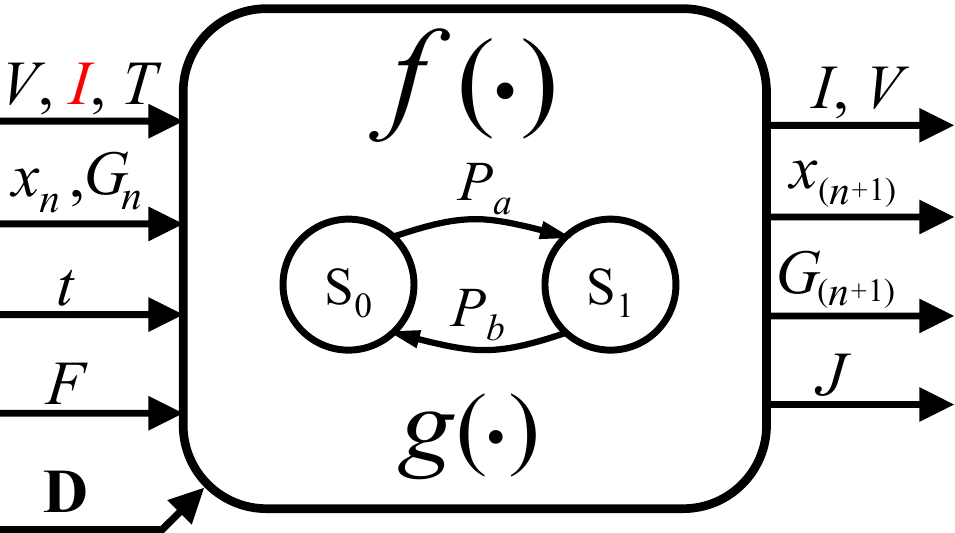}
 \caption{\small A generalized form of modeling in-situ stochastic behavior of a given device (in this paper, RRAM). Inputs consist of applied voltage, $V$ (or current, $I$), operating temperature, $T$, signal pulse-width, $t$, initial state $x_n$ and initial conductance $G_n$. Device type and relevant fitting parameters are given by $\mathbf{D}$. The model accepts cross-point yield as a defect rate input $F$ (not to be confused with the material\rq{}s point defect). Function $f(\cdot)$ defines switching probability and function $g(\cdot)$ defines conductance value randomness under a given C2C variation distribution. Switching probabilities are defined as $P_a$ and $P_b$ for switching from logic state `0\rq{}, S$_{\rm 0}$, to logic state `1\rq{}, S$_{\rm 1}$, and from logic state `1\rq{} to logic state `0\rq{}, respectively. Outputs are current and voltage responses, next state ($x_{(n+1)}$), next conductance ($G_{(n+1)}$), as well as total consumed energy $J$ of the whole operation to performing a programming or sensing, regardless of its success.}\label{fig:func}
\end{figure}

Stochastic switching of a nanodevice could be treated as a black-box; a generalized model, which is identified by a set of inputs and outputs. Fig.~\ref{fig:func} provides an overview of the model. Outputs $x_{(n+1)}$ and $G_{(n+1)}$ represent next state and next conductance of the device, which both are probabilistic and their values are defined by the functions $f(\cdot)$ and $g(\cdot)$, respectively. Once $x_{(n+1)}$ is determined, $G_{(n+1)}$ is calculated using an experimentally verified distribution (in this paper, log-normal) of ON and OFF state resistances. The function $f(\cdot)$ is a function of voltage/current, time (here pulse-width) and operating temperature. Small updates in $g(\cdot)$ is independent of function $f(\cdot)$ output, which means if $x_{n}=x_{(n+1)}$, $G_{n}$ and $G_{(n+1)}$ may not be equal.    

\subsection{Probabilistic plasticity}\label{subsec:stoch-plastic}
\begin{figure}[!ht]
\centering
  \includegraphics[width=0.70\textwidth]{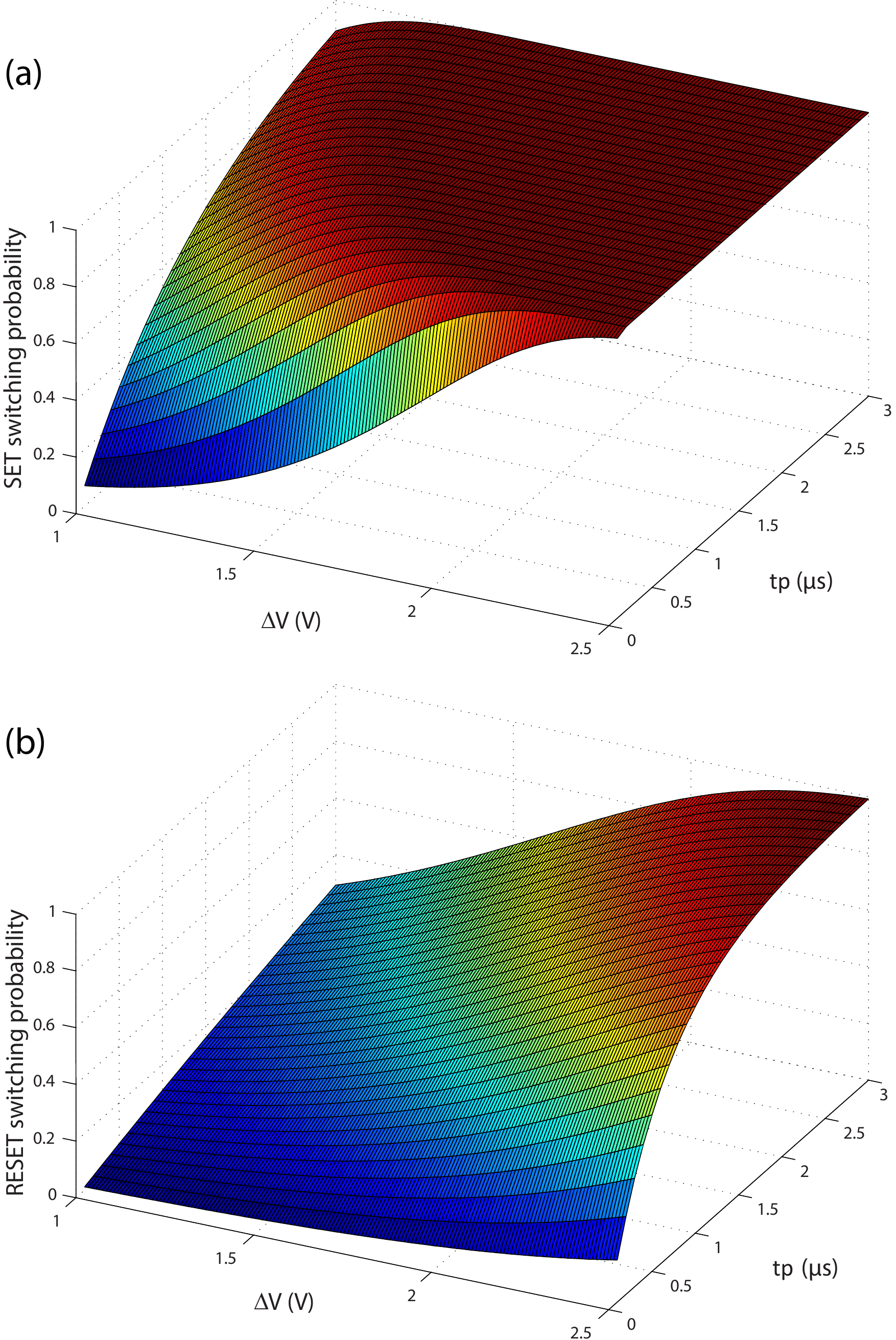}
 \caption{\small Output of function $f(\cdot)$. (a) demonstrates SET switching probability and (b) shows RESET switching probability. $t_{\rm p}$ represents applied voltage pulse-width.}\label{fig_swprob}
\end{figure}

Experimental data on voltage, pulse-width and temperature dependency of a resistive memory systems is gathered from \cite{siddharth2013stochastic,suri2012cbram,kavehei2011fabrication,kavehei2013associative,kuzum2013synaptic,park2013analysis,chang2013high,yu2013low}. Device switching probability is defined as a strong function of applied voltage, $\Delta V$, and time, $\Delta t$. For simplicity, temperature dependency and switching threshold variation of the device are considered to have an overall impact on the switching probability, therefore, their impact are not discussed as independent variables. Fig.~\ref{fig_swprob} demonstrates switching probability of the SET (S$_0\rightarrow$S$_1$) and RESET (S$_1\rightarrow$S$_0$). Under similar condition, a SET is more likely than a RESET. Given that the S$_0\rightarrow$S$_1$ seems more likely, it is possible to modulate the probabilistic LTP using a complicated learning rule, $\Delta W$, where $W$ represents synaptic weight that is bounded between a maximum and a minimum limits, through a different programming pulse-scheme for LTP, which makes the design more complex. Experimentally verified information rgarding the switching probability model are given in \cite{siddharth2013stochastic}. The switching probability is achieved by integrating a Poisson-like distribution, which is in an excellent agreement with the hypothesis of thermodynamic activation during filament formation over a dominant energy barrier in RRAM devices. 

The mentioned experimental studies suggest that the switching probability of redox-based memory is a cumulative phenomenon, which links the digital and analog characteristic of the device. The probabilistic SET and RESET, under such circumstance, can therefore be shown as a function of number of applied pulses with a given amplitude-time combination to model the history dependent effects. One could argue that such spike-history-dependent-plasticity provides further phenomenological similarities between these artificial synapses and the chemical synapses, where history dependent effects also play an important role \cite{manwani1998synaptic}. 

\begin{figure}[!ht]
\centering
  \includegraphics[width=0.90\textwidth]{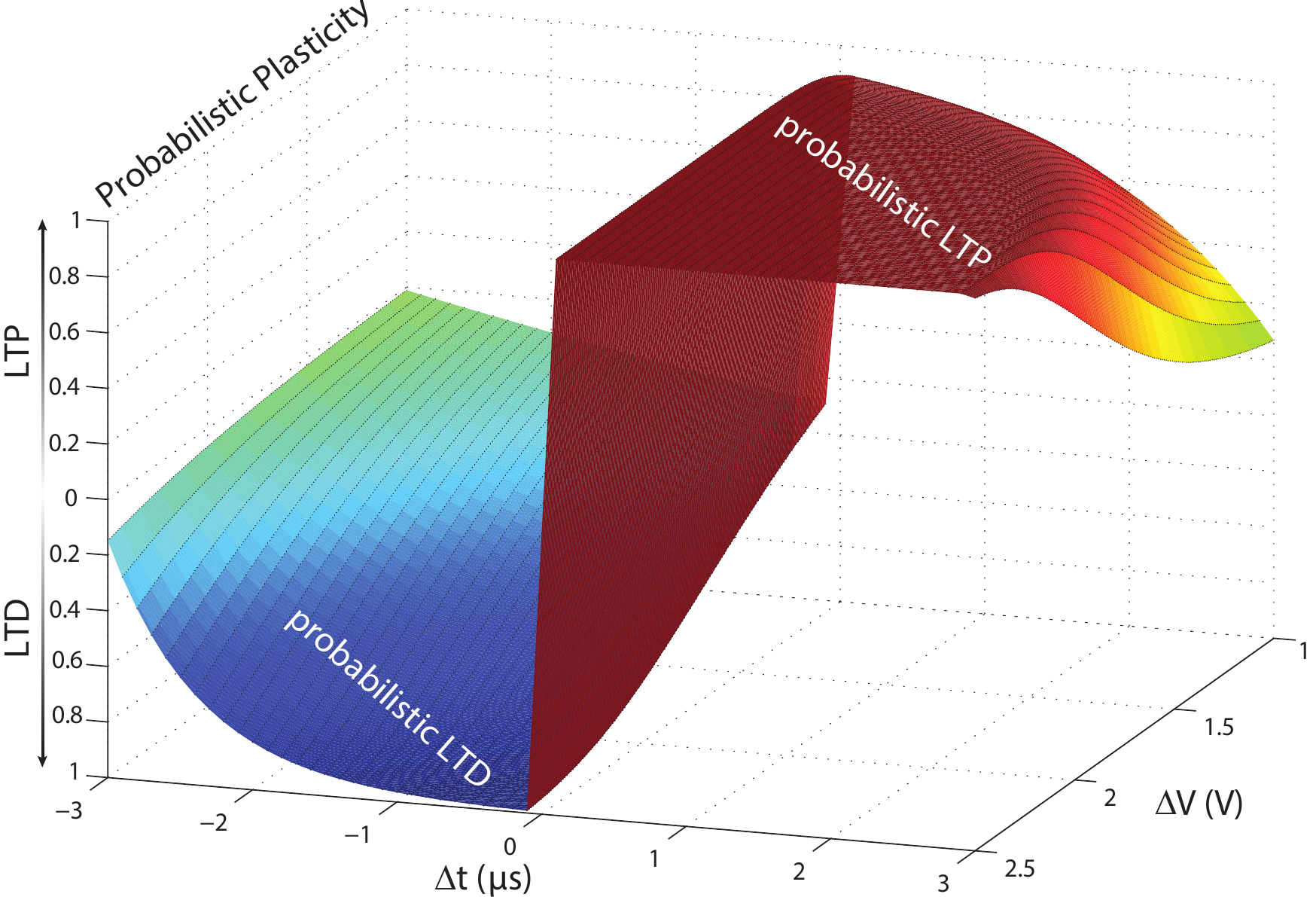}
 \caption{\small Probabilistic plasticity that is `solely\rq{} implemented using the in-situ stochastic switching of a 1-bit synapse.}\label{fig_plasticity}
\end{figure}

This stochastic behavior in combination with $\Delta W$ results a probabilistic LTP (long-term potentiation) and LTD (long-term depression), which is a function of time difference ($\Delta t$) between a pre- and a post-synaptic spike (probabilistic spike timing-dependent plasticity, pSTDP) or their difference in voltage ($\Delta V$). Stochastic change of the device state can also occur under a sequence of applied pulses with a fixed $t_p$. The probabilistic plasticity, which defines the probability of change in the device state, is shown in Fig.~\ref{fig_plasticity}. Here $\Delta W$ is defined as

\begin{eqnarray}
\Delta W = \begin{cases} 1, & \mbox{if } \Delta t>0~, \\ -1, & \mbox{if } \Delta t<0~. \end{cases}\label{equ:deltaw}
\end{eqnarray} 

\begin{figure}[!ht]
\centering
  \includegraphics[width=0.95\textwidth]{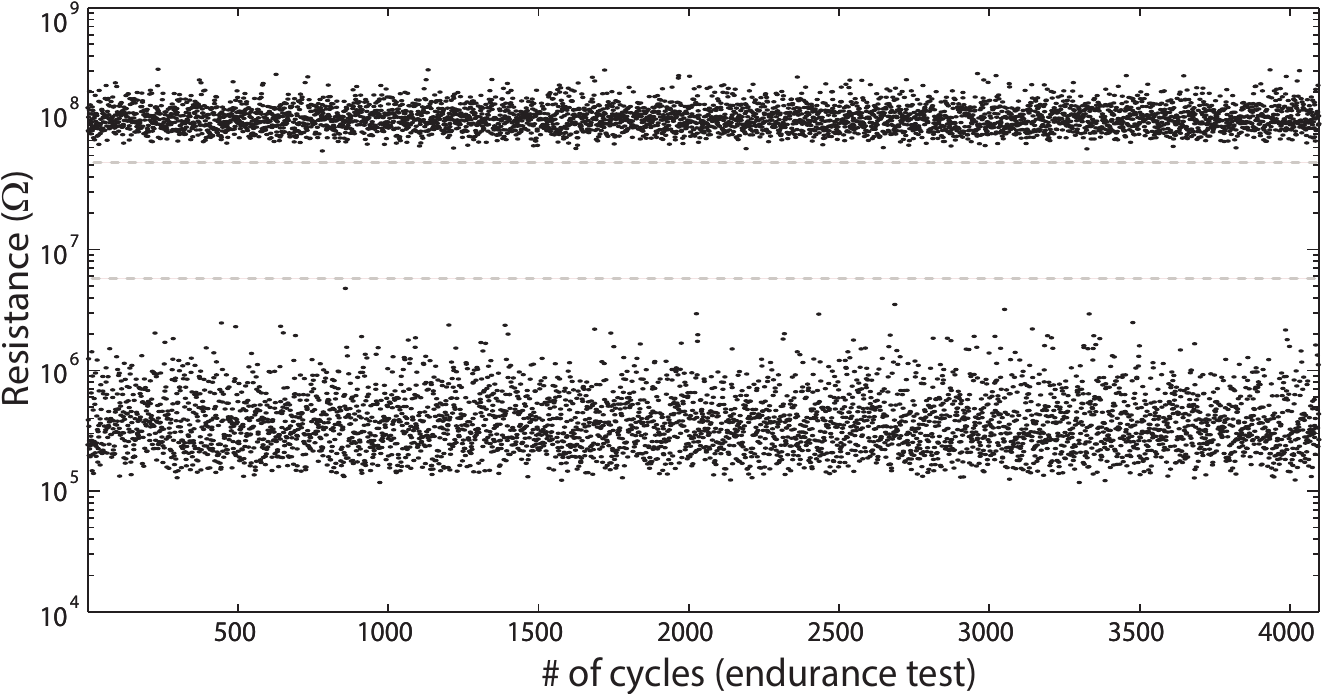}
 \caption{\small OFF and ON resistance variation of a single RRAM for $>$\,$4000$ switching cycles. This behavior is captured by $g(\cdot)$.}\label{fig_res}
\end{figure}

It can be shown that a simple form of $\Delta W$, like Eq.~(\ref{equ:deltaw}), is enough to successfully implement a robust and reliable selectivity task. There is a significant cycle-to-cycle resistance variation. This uncertainty is defined and verified using the function $g(\cdot)$. The RRAM and CBRAM devices have experimentally shown log-normal distribution of ON and OFF state resistances. Fig.~\ref{fig_res} illustrates high and low resistance variation of a RRAM device for more than $4000$ switching cycles. In some devices, like \cite{suri2012cbram}, OFF state resistances may shows wider variation due to the uncontrolled metallic filament dissolution. 

\section{Orientation Sensitivity}\label{sec:select}
\begin{figure}[!ht]
\centering
  \includegraphics[width=0.90\textwidth]{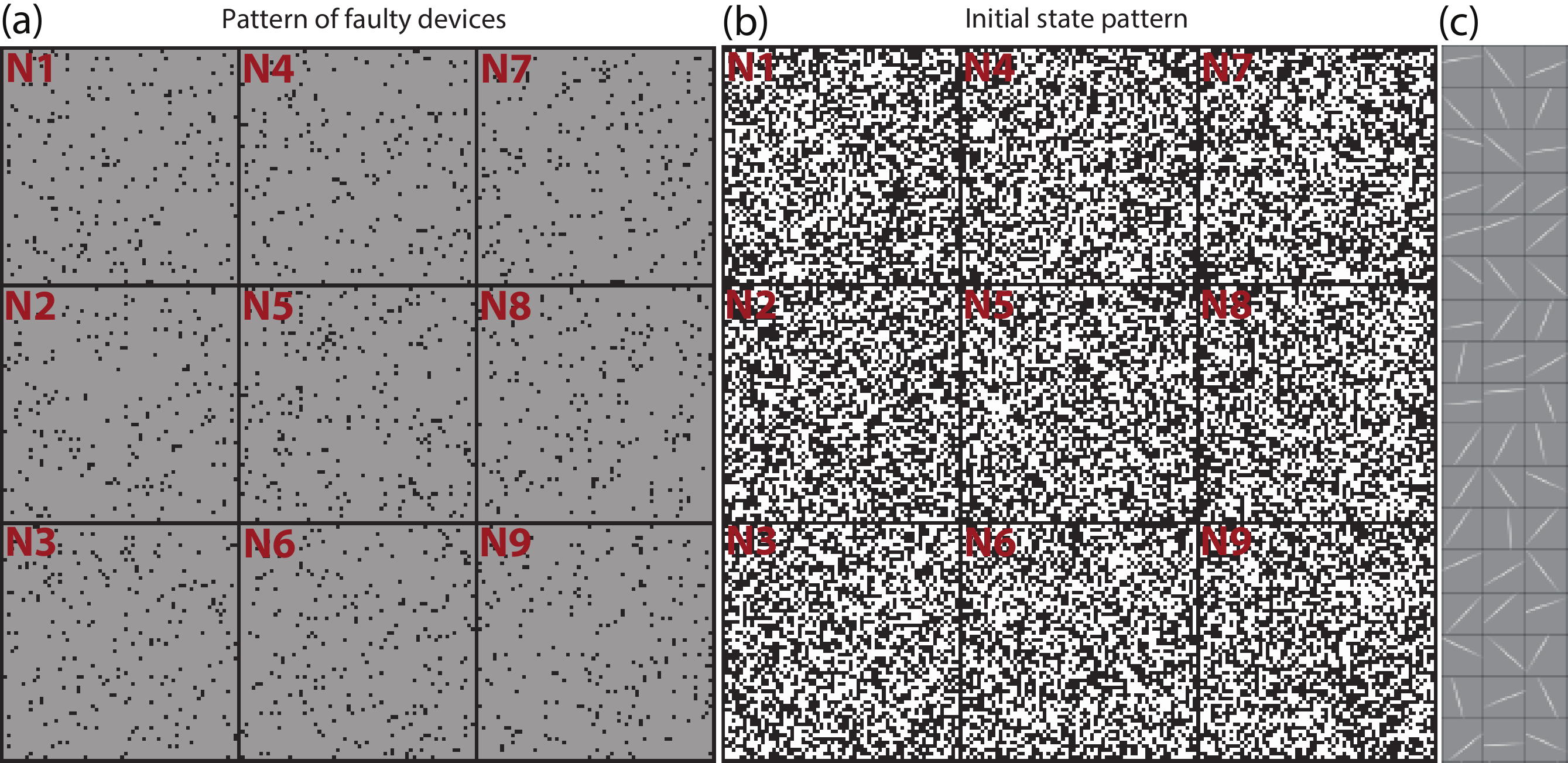}
 \caption{\small (a) shows device defect patterns. A black dot represents a faulty device, which is not able to switch (either stuck-at-ON or -at-OFF). Yield is $95\%$ and $70\%$ of faulty devices are stuck-at-ON. (b) demonstrates initial random patterns of ON and OFF devices. Black pixels show OFF state devices and white pixels are ON device. Half of the functioning devices are randomly programmed as ON, and the remaining as OFF. (c) some of the randomly generated input patterns are shown here.}\label{fig_result1}
\end{figure}

A two layer neural network is implemented with $64\times 64$-pixel binary visual images. Inputs were generated using a Gabor-like function to produce more than a thousand, randomly centered, receptive fields. Each input pixel drives an input neuron that spikes if the pixel represents `1\rq{}. There are nine output neurons that are fully connected to all input neurons through one-bit resolution stochastic RRAM connections. Therefore, $36,864$ synapses are available that are randomly programed and some are faulty. Faulty device patterns, random initial state, and some of the visual inputs are shown in Fig.~\ref{fig_result1}. Output neurons are integration-and-fire neurons that are connected to each other through inhibitory connections, which form a winner-takes-all network. Output neurons could be configured to implement Hebbian, anti-Hebbian, STDP, and anti-STDP learning with a support for a programmable refractory period. The system implements an unsupervised competitive learning that is capable of accepting a wide range of input patterns, including visual and auditory with or without preprocessing.

Energy per spike is an important factor. The model is able to estimate both dynamic switching energy and static leakage power through unselected devices and sneak current paths with or without selector devices. Although, there is little experimental information available regarding spatial variation (device-to-device), the model aims to consider overall impact of such variation with some worst-case assumptions.

\begin{figure}[!ht]
\centering
  \includegraphics[width=0.90\textwidth]{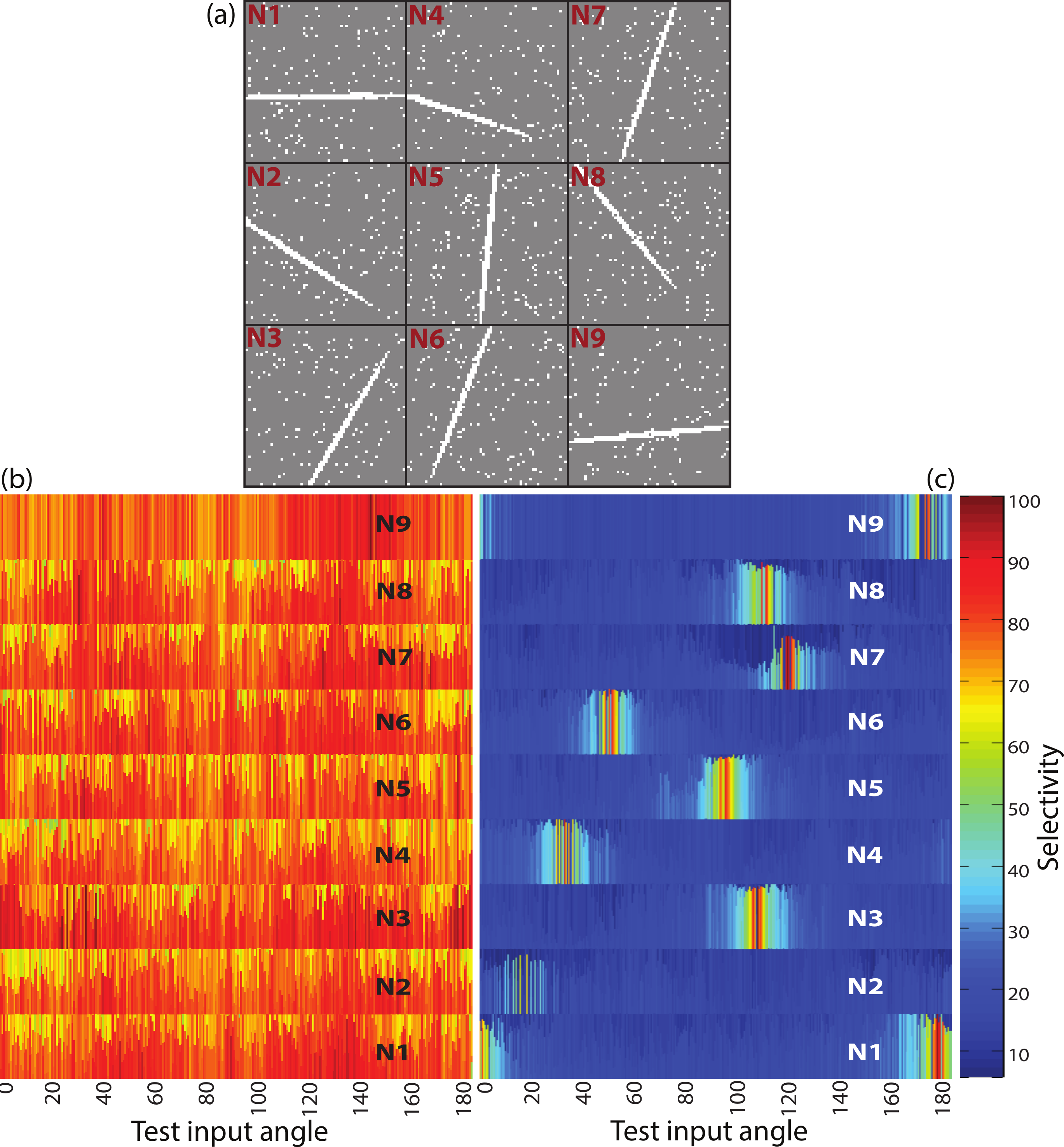}
 \caption{\small (a) output neurons after training. Random white pixels are either faulty stuck-at-ON devices or ON state devices. Output neurons\rq{} sensitivity to $>$\,$10,000$ test patterns before training (in response to Fig.~\ref{fig_result1}b) and after training are shown in (b) and (c), respectively.}\label{fig_result2}
\end{figure}

During training, inputs were shown to the network and each was repeated several times. After training, as shown in Fig.~\ref{fig_result2}a, output neurons become sensitive to several orientations. Fig.~\ref{fig_result2}b shows that the system was not sensitive to any angle prior to learning. The system sensitivity to particular angles of the visual input patterns is demonstrated in Fig.~\ref{fig_result2}c. One of the main parameters of designing the system is the dynamic tuning of the output neurons\rq{} firing threshold, which is not the focus of this paper. The training process were repeated with different threshold tuning approaches and a range of initial conditions, device variation, and defect patterns. The system continues to perform reliably under extreme yield (defect rate is $>$\,$20\%$) and variability assumptions, where the function $f(\cdot)$ output varies within $0.1$ probability ($3\sigma$) around its nominal value. Implementing such system with low-power synapses would produce an ultra-high energy efficient and highly scalable system with an energy consumption of $<$\,$1$~pJ/spike.

\section{Conclusion}\label{sec:conc}

The implemented neuromorphic system is (i) highly scalable, (ii) easily mappable on the current crossbar memory architecture, (iii) not sensitive to device variation, (iv) working reliably with or without selector devices, (v) potentially very high energy efficient, and (vi) potentials for high throughput applications. Each training process, in this case, demands for an endurance requirement in the order of $>$\,$20$k cycles, which is well within the current reported endurances of RRAM, CBRAM, PCM, STT-MRAM, MeRAM, and atomic switches. Cost-wise, implementing such network with RRAM, CBRAM, PCM, and atomic switches may be cheaper relative to STT-MRAM and MeRAM. However, rapid progress of these technologies promises high energy efficient memory systems, which makes the cost-energy trade-off of these devices more attractive in the near future. Modeling accuracy can be further improved through systematic studies like \cite{siddharth2013stochastic}.

\section*{Acknowledgements}
Omid Kavehei would like to thank Dr Shimeng Yu for his insightful discussions on memory technologies and Ms Nafise Erfanian Saeedi for her comments on the neural network implementation and selectivity.

\bibliographystyle{IEEEtran}

\end{document}